# Mechanisms of superconductivity and inhomogeneous states in metallic hydrogen and electron systems with attraction

*M. Yu. Kagan (MIEM HSE, Kapitza Institute) A.V. Krasavin (NRNU MEPHI, MIEM HSE), R. Sh. Ikhsanov (MIEM HSE, Lebedev Institute), E.A. Mazur (NRNU MEPHI, Kurchatov Institute), A.P. Menushenkov (NRNU MEPHI)*

**Abstract**

In the Review we discuss anomalous aspects of superconductivity (SC) and normal state, as well as formation of inhomogeneous (droplet-like or cluster-like) states in electron systems with attraction. We consider both the models with the retardation (Eliashberg mechanism of SC for strong electron-phonon interaction in metallic hydrogen) and without retardation (but with local onsite attraction). We concentrate on the mechanism of the BCS-BEC crossover for the Hubbard model with local attraction and diagonal disorder for the two-dimensional films of the dirty metal. In 2D Hubbard model in the framework of the Bogoliubov-De Gennes (BdG) approximation for strong interaction and strong diagonal disorder at low electron densities the inhomogeneous states are realized in the system with the droplets of the order parameter in the matrix of unpaired states as well as the percolating insulator-superconductor phase transition when we increase electron density. We analyze also the model of the inhomogeneous space-separated Fermi-Bose mixture for the bismuth oxides BaKBiO, which contains the paired clusters of bosonic states as well as unpaired fermionic clusters. This model explains the unconventional phase diagram of the system containing the anomalous phases of bosonic insulator, bosonic semiconductor and bosonic metal. Superconductivity is realized in this system due to local pairs tunneling from one bosonic cluster to the neighboring one via the fermionic barrier. For metallic hydrogen and metallic hydrides, we calculate the critical temperature and discuss important possibility for practical applications how to increase the temperature by decreasing pressure in the framework of the generalized Eliashberg approach. We advocate also interesting analogies with the quantum (vortex) crystal for long-living low-dimensional metastable phases of metallic hydrogen including filamentous phase with proton chains embedded in 3D electron Fermi liquid and planar phase with proton plains. We formulate the concept of two Bose-condensates in SC electron and superfluid (SF) ion subsystems and provide the estimate for the lifetime of the long-living metastable phases at normal pressure. The estimate is connected with the formation and growth of the critical seeds of the new (molecular) phase in the process of quantum under-barrier tunneling.

## 1. Introduction. Local and extended pairs. BCS-BEC crossover in quantum gases and attractive-U Hubbard model.

As we know, Cooper pairs in traditional metals are extended and strongly overlapping in the real space [1-2]. SC pairing in the BCS theory [1-3] takes place in momentum space against the background of the filled Fermi sphere. SC transition is determined by one critical temperature $T_C$ (Cooper pairs are formed and Bose-condensed at the same temperature). At the same time in a number of systems an emergence of the local pairs is possible [4-7]. Local pairs appear in quantum Fermi gases of $Li_6$ and $K_{40}$, in the restricted geometry of magnetic and dipole traps [8, 9], in SC bismuth oxides BaKBiO [10-12], and (possibly) in under-doped high-$T_C$ cuprates [13-14]. SC pairing takes place in the coordinate space [4-7, 15]. In this case in three-dimensional systems in the domain of Bose-Einstein condensation (BEC domain) [16, 17] appear two characteristic temperatures. The higher temperature $T^* \sim |E_b|$ is related to the formation of local pairs and is determined by the Saha formula [18, 19] for the dynamic equilibrium of paired and unpaired particles ($|E_b|$ is the binding energy of the local pair). This temperature corresponds to the smooth crossover. At the same time the second (critical) temperature $T_c \sim 0.2\varepsilon_F$ [6-7] in the main approximation is determined by Einstein formula and is related to the real phase transition ($\varepsilon_F$ is the Fermi energy). For the intermediate temperatures $T_C<T<T^*$ an interesting new phase of bosonic metal is realized [20-25].

On Fig.1 we present the typical phase diagram of the BCS-BEC crossover in the basic model of the 3D Fermi gas with attraction [26]. The phase diagram is constructed in the axes of the dimensionless temperature $T/\varepsilon_F$ and inverse gas parameter $1/ap_F$ ($a$ is the scattering length, $p_F$ is Fermi momentum). We show BCS and BEC domains on the phase diagram. The scattering length $a$ is negative in rarified BCS domain (for $a<0$ and $|a|p_F \ll 1$) and is positive in rarified BEC domain ($a>0$ and $ap_F \ll 1$) due to the formation of the bound state with the energy $|E_b|$. The chemical potential in the rarified BCS-domain is positive ($\mu > 0$) and approximately equal to the Fermi energy: $\mu \sim \varepsilon_F$. At the same time, in the rarified BEC domain the chemical potential is large and negative: $\mu \sim -|E_b|/2 < 0$ when $|E_b| \gg \varepsilon_F$.

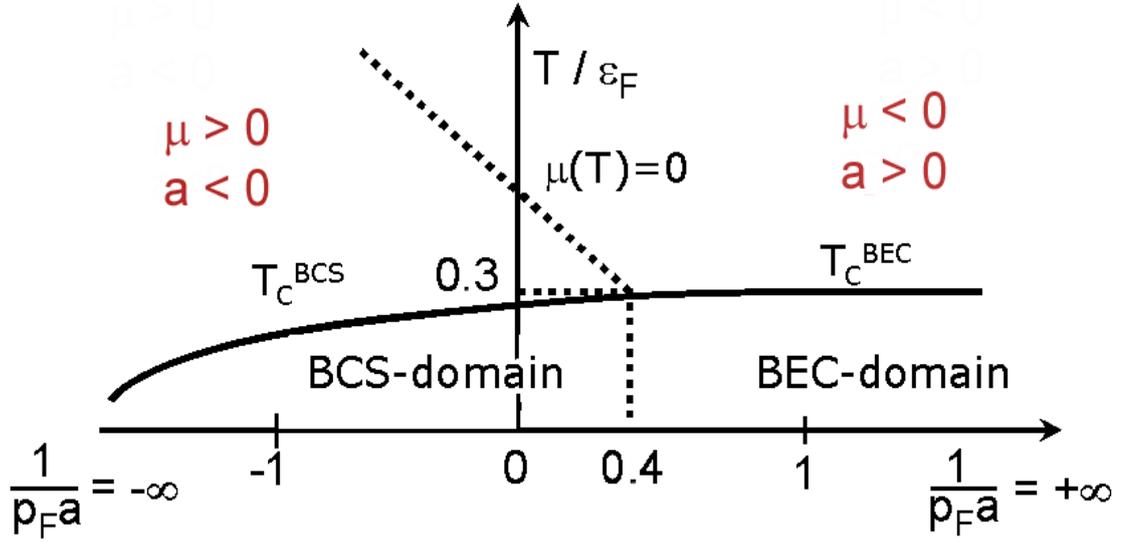

**Fig. 1.** Typical phase diagram of the BCS-BEC crossover in the 3D Fermi gas with attraction. The phase diagram is presented in the axes of the dimensionless temperature $T/\varepsilon_F$ and inverse gas parameter $1/ap_F$. On the figure $a$ is the scattering length, $\mu$ is the chemical potential [26].

## 2. Droplets of the order parameter in the two-dimensional Hubbard model with strong attraction in the presence of strong diagonal disorder

Two-dimensional Hubbard model with short range (onsite) attraction between electrons and diagonal disorder has the following form on the square lattice [27-33]:

$$\tilde{H} = H - \mu N = -t \sum_{ij\sigma}(c^+_{i\sigma} c_{j\sigma} + h.c.) - |U|\sum_i n_{i\uparrow}n_{i\downarrow} + \sum_{i,\sigma}(V_i - \mu)n_{i\sigma}, \quad (1)$$

where $|U|$ is the absolute value of Hubbard onsite attraction, $t$ is the amplitude of hoping on the neighboring site, $c^+_{i\sigma}$ and $c_{j\sigma}$ are creation and annihilation operators on site $i$ with spin projection $\sigma$ correspondingly, $n_{i\sigma}$ is the local electron density on site $i$ for one spin projection, $\mu$ is the chemical potential. The random potential $V_i$ on site $i$ describes diagonal (local) disorder and is homogeneously distributed in the interval [-V, V]. We consider 2D Hubbard model for s-wave disordered SC in the limit of strong local attraction ($|U|>W=8t$) and strong diagonal disorder ($V>W$) [27-29].

At small electron densities ($n = 2\varepsilon_F / W \ll 1$) in the framework of the Bogoliubov-De Gennes (BdG) approximation [34-35] for the attractive-$U$ Hubbard model [31-33] we get an interesting result with the appearance of the order parameter droplets in the matrix of unpaired states [27-30]. When we increase electron density till the value of $n$=0.31, the SC droplets start to form large percolation cluster and thus insulator-superconductor phase transition takes place in the system [27-29]. The numerical results or the formation of the inhomogeneous droplet structures are presented on Fig. 2. We show space distribution of local electron density (left column), formation of the order parameter droplets (right column) and electron-hole mixing typical for BdG approximation (middle column).

For strong Hubbard attraction $|U|>W$ and low electron densities $n \ll 1$, our system demonstrates the features of the bosonic metal. The concept of this new metallic state was introduced in [20-24] for the 2D Hubbard model with sufficiently strong Hubbard attraction and small electron density in the clean case (in the absence of impurities and disorder) in [20-21]. In this new metallic state both in SC and normal regions, the compact (local) electron pairs serve as charge carriers, which can form more extended bosonic clusters (see e.g. results of [10-12] and the next section of the Review on bismuth oxides). In our case bosonic clusters can contain one, two or several compact (bi-electron) pairs. More rigorously, in this case we will have space-separated Fermi-Bose mixture of compact Cooper pairs and unpaired electrons with the formation of bosonic droplets of different size in the matrix of unpaired fermionic states [27-29].

Percolating phase transition for density close to *n*=0.3

Let us concentrate now on the region of electron densities close to the value of *n*=0.3 in the limit of strong interaction and strong disorder $|U| \sim V > W = 8t$. In this range of parameters as it was shown in [27-29], the order parameter droplets merge into a network of paired chains and then the network of paired chains forms a large percolation cluster resembling a tree with a lot of branches, which practically do not exhibit discontinuities along the lengths. As a result, we have a percolation phase transition from granular to SC state. Preliminary estimates for the percolation transition yield the critical concentration $n_C \approx 0.31$ [27-29].

Discussion.

Let us estimate the size of the electron pair in the limit of small density and strong coupling. If in the absence of disorder (for $V/t$ =0) and low electron density (e. g. for *n*=0.125) in the strong coupling limit $|U|/t$=10 we make simple calculations (see [8-10, 16, 23-24]), then we get $E_F \sim 0.8t$ for the Fermi energy and $|E_b| \sim 0.2\, E_F \sim 0.16\, t$ for the binding energy of the pair. If we further decrease the density till the limiting value *n*=0.05 (for numerical calculations), fixing the strength of the Hubbard attraction, then we get $E_F \sim 0.32\, t$, and $|E_b| \sim 0.5$, $E_F \sim 0.16\, t$.
Thus, even in such extremely low-density limit:

$$|E_b| < 2\varepsilon_F, \qquad (2)$$

and the chemical potential is still positive:

$$\mu = \varepsilon_F - \frac{|E_b|}{2} = 0.75\, \varepsilon_F > 0. \qquad (3)$$

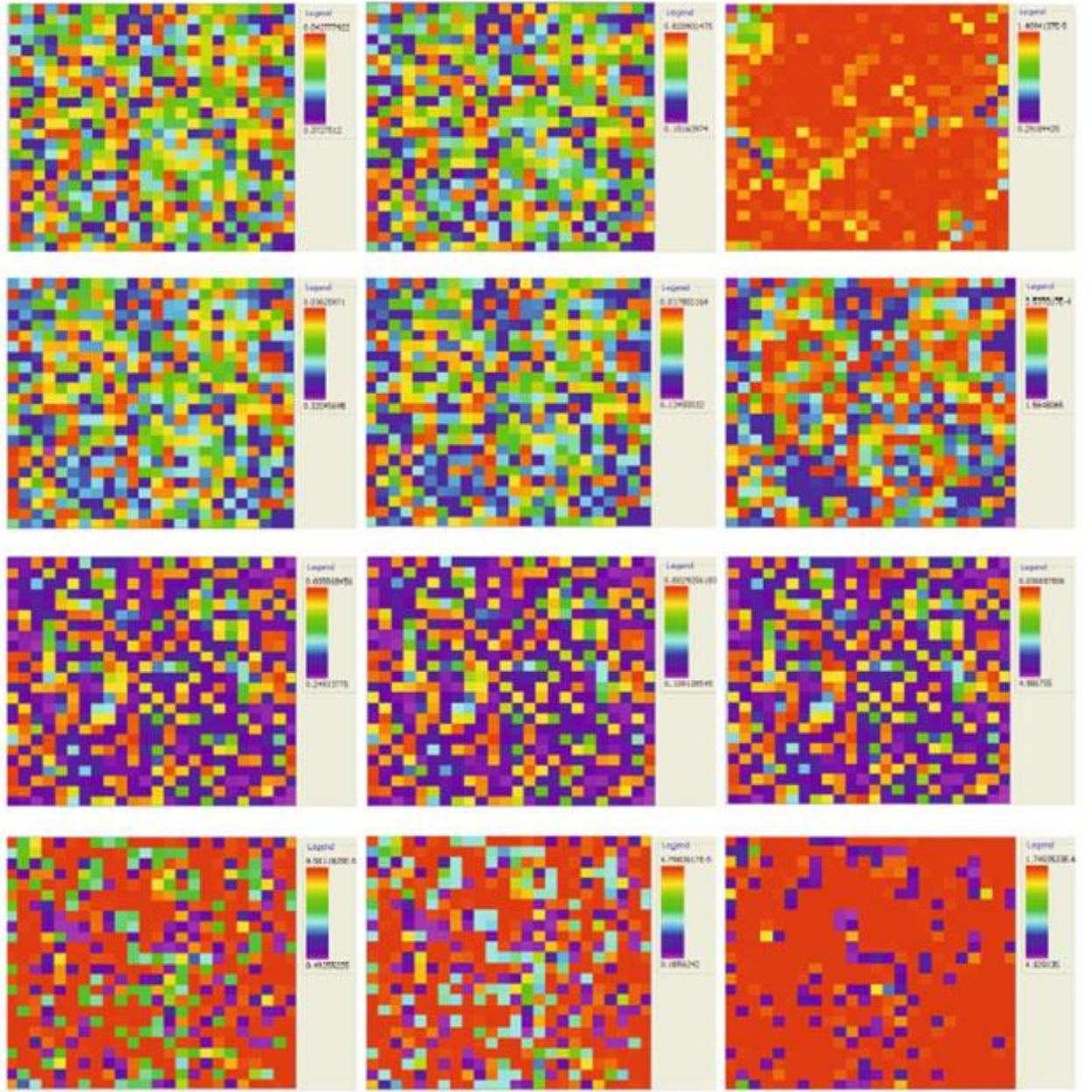

**Fig. 2.** Spatial distribution of the local electron density (left column), order parameter droplets (right column), and mixing of holes and electrons (middle column) in four different cases, namely: ***a*** for weak Hubbard interaction and weak disorder, ***b*** for moderate Hubbard interaction and weak disorder, ***c*** for strong Hubbard interaction and weak disorder, and ***d*** for strong Hubbard interaction and strong disorder. The averaged electron density *n*=0.15 is fixed in all the cases [27].

Hence, even in this case the pairing takes place close to the Fermi surface and we are still at the BCS side of the BCS-BEC crossover [6-7]. Note, that if we introduce convenient notation for the binding energy

$$|E_b| = 1/ma^2,  \qquad (4)$$

where $a$ is the effective size of the pair, $m = 1/2td^2$ is the band mass, $d$ is intersite distance, $\varepsilon_F = p_F^2/2m$ is Fermi energy, and $p_F$ is Fermi momentum, then the inequality (4) means, that $ap_F$, an important parameter of the model [23-24, 27], is subject to inequality

$$ap_F = \left(\frac{2\varepsilon_F}{|E_b|}\right)^{1/2} > 1, \qquad (5)$$

and the size of the pair $a$ exceeds the mean distance between electrons $1/p_F$.

Let us emphasize that the size of the pair $a$ in Eq. (5) qualitatively resembles the coherence length. At the same time, for extremely low density $n=0.05$, we have $ap_F \sim 2$, which is much less than the corresponding parameter $p_F \xi_0 \sim 1000$ in conventional low-$T_C$ $s$-wave superconductors, where the typical coherence length $\xi_0 \sim (3 \div 5) \times 10^3$ Å [1-3]. Thus, we can conclude that though we are at the BCS side of the crossover, the pairs are rather compact and almost touch each other. In fact, we are very close to the limit when the pairs start to overlap and "crush" each other [23-24]. This limit effectively corresponds just to the formation of the Fermi-Bose mixture [10-12, 36-37] with the coexistence of compact pairs and unpaired single electrons. The presence of disorder makes the situation even more interesting and complicated and leads to the formation of the inhomogeneous Fermi-Bose mixture of pairs and single electrons, where a part of the one-particle and two-particle states should be localized even in the 2D system of a finite size depending upon the degree of disorder $V/t$ [27].

Note, that important feature of the quasi-2D systems, described by the Hubbard-Anderson type of models [36] with local onsite Hubbard interaction and diagonal disorder is related to the fact that their global phase diagrams can exhibit a "direct" transition from the SC to insulating localized state in addition to the conventional phase transition from SC to normal metal. This possibility is elucidated in theoretical papers [38-41] and confirmed by the experimental studies of superconductivity and localization in dirty thin films [41-43]. Another important motivation for our study is related to recent experimental efforts to create SC flux qubits using quantum circuits with high impedance in granular superconductors of reduced dimensionality [44]. These motivations promote the calculations presented in this section and make them highly relevant,

## 3. Local pairs formation and superconductivity in bismuth oxides Ba K Bi O

In papers [10-12] we constructed phase diagrams of SC and normal states in bismuth oxides $Ba_{1-x}K_xBiO_3$ for different potassium concentrations $x$. For $x=0$ the parent compound $BaBiO_3$ can be described by the anomalous state of bosonic isolator. This state corresponds to the formation of local bi-electron pairs in $BiO_6$ clusters in agreement with the predictions of Anderson theory of negative-$U$ centers [45] and Varma ideas [46] on valence disproportionation in bismuth oxides. Due to the ideas of Varma each $BiO_6$ cluster can contain 18 or 20 electrons. Thus, the number of electrons on one cluster can differ on $2e$, that is precisely on the local pair. At the same time, formation of negative-$U$ centers (which leads to attraction) presumably has a quantum-chemical nature in bismuth oxides. Note that emergence of local pairs in the parent compound $BaBiO_3$ was confirmed recently by Menushenkov et al. [12] in experiments on X-ray absorption spectroscopy (XAS) with femtosecond resolution. Experiments were performed on the European X-ray free-electron laser (EuXFEL) in Schonefeld (Germany) in 2024. The authors of [12] observed the resonance destruction of local pairs by optical laser pulse on the laser frequency which equals to the pair binding energy $\omega = E_b$.

Let us stress that for potassium concentrations different from zero the space-separated Fermi-Bose mixture of bosonic clusters (containing local pairs) and fermionic clusters (related to unpaired states) starts to form in the system [10-12]. At small concentrations fermionic and bosonic subsystems are also separated in the energy space. As a result, for $T=0$ the system remains insulating [12]. However, for concentrations $x=0.37$ the percolation phase transition takes place, and thus, the bismuth oxide $Ba_{1-x}K_xBiO_3$ becomes metallic and superconducting at low temperatures (see Fig. 3). Note that metallic state here is also very peculiar and can be specified as bosonic metal which is shunted by the fermionic component. In this state fermionic and bosonic subsystems are still separated in the real space, but their separation in the energy space disappears [10-12]. Finally, superconductivity in this system is realized due to local pairs tunneling from one bosonic cluster to the neighboring one via the fermionic barrier.

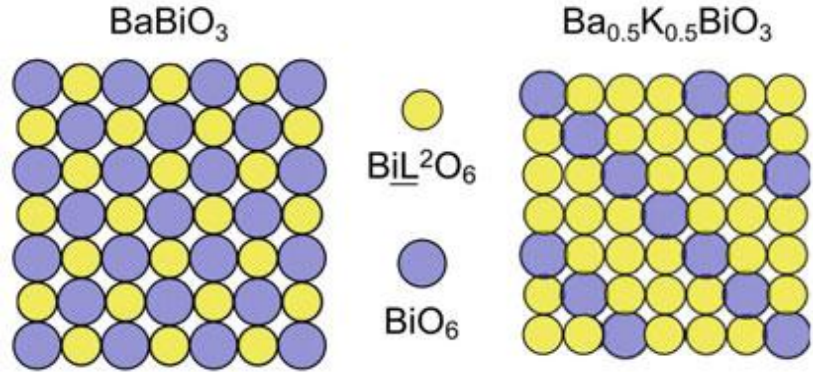

**Fig. 3.** Local crystal structure of the parent compound (bosonic isolator) $BaBiO_3$ with the checker-board distribution of $BiO_6$ octahedra with 18 and 20 electrons (left Figure). On the right Figure we show formation of large percolation cluster in SC bosonic metal $Ba_{0.5}K_{0.5}BiO_3$ [11].

### 3. Eliashberg mechanism for metallic hydrogen and hydrides of metals

For metallic hydrogen [47-48] and most of double and triple metallic hydrides, such as $H_2S$, $H_3S$, $LaH_{10}$ and others [49-51], the basic mechanism of SC is determined by the Eliashberg theory [52] for strong electron-phonon interaction. Calculations based on Eliashberg theory [53-56], predict SC for experimentally detected hexagonal phase of metallic hydrogen stabilized at pressure $P=5$ MBar under the critical temperature in the range of (215÷217) K.

These calculations confirm the early ideas advocated by Ashcroft already in 1968, that metallic hydrogen due to its light nucleus mass (actually the mass of proton) and large Debye frequency can serve as a very promising candidature for the realization of room-temperature superconductivity [57]. Note that the main efforts of physicists, chemists and material scientists today shifted from the problem to increase the critical temperature to the problem of the pressure decrease and search of the long-living metastable phases of metallic hydrogen and hydrides of metals, which are SC at normal pressure. It was demonstrated that the complication of the compound composition in metallic hydrides, and in particular the shift in the investigations from double to triple hydrides [58], promotes effective pressure decrease conserving at the same time high critical temperatures.

Search for the long-living metastable phases of metallic hydrogen

In this context we find very promising the early predictions of Yu. M. Kagan, E.G. Brovman and A. Holas [59-60], dated back to nineteen seventies. According to the results of [59-60], several low-dimensional phases of metallic hydrogen and in particular quasi one-dimensional filamentous phase (with proton chains embedded in 3D electron Fermi liquid) can be stabilized not only at high pressure of the order of several megabars but can be experimentally realized also as a long-living metastable phase at lower pressures and maybe even at atmospheric pressure.

Note that the decay of the metastable metallic phase is fulfilled via the virtual formation and growth of the critical seed of the stable molecular phase in the process of quasi-classical under-barrier tunneling in the effective anharmonic potential. This anharmonic potential in agreement with the theory of I. M. Lifshitz and Yu. M. Kagan [61] is determined by the sum of the surface contribution (proportional to $R^2$) and bulk contribution (proportional to $R^3$) (see Fig.4):

$$U(R) = 4\pi\sigma R^2 - \frac{4\pi R^3 n_2(\mu_1-\mu_2)}{3}, \tag{6}$$

where $R$ is the radius of the spherical seed of molecular phase, $\sigma$ is surface tension coefficient, $\mu_1$ and $\mu_2$ are the chemical potentials of metallic and molecular phase, $n_2$ is the density of molecular phase.

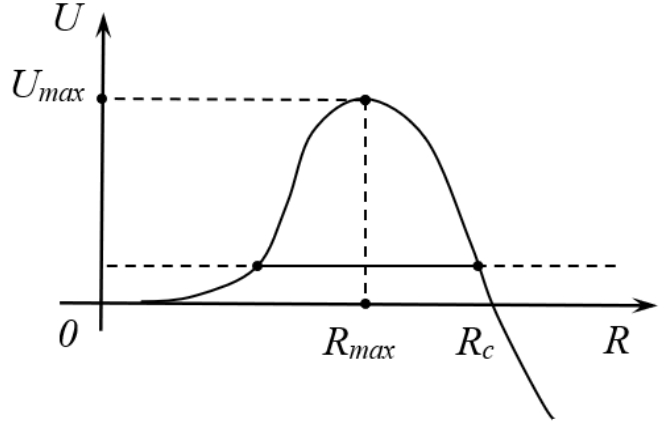

**Fig. 4.** Metastable (metallic) and stable (molecular) state of metallic hydrogen serving as local and global extrema of effective interaction. We show on the figure quasi-classical under-barrier tunneling of the seed of the molecular phase in effective anharmonic potential which is determined by the sum of surface contributions (proportional to $R^2$) and bulk contribution (proportional to $R^3$), where $R$ is the radius of the spherical seed of molecular phase [61].

Recent theoretical estimates of Burmistrov and Dubovskii [62] developing ideas of [59-60] demonstrate in particular that the filamentous phase the region of long-living metastable states is extended at least up to the pressures of the order of $P \sim 0.1$ Mbar.

Possible analogy with quantum (vortex) crystal

Note, that low-dimensional phases of metallic hydrogen and first of all filamentous phase (with proton chains) and planar phase (with proton planes) have many similarities with quantum crystals. In particular they have a lot in common with the physics of super-solidity in quantum helium crystals and the ideas of the classical paper by A.F. Andreev and I.M. Lifshitz. [63]. The similarities first of all are related to the values of De Boer and Lindemann melting parameters in the quantum regime [23, 25, 64]. In papers [51, 64] we formulated qualitative considerations about possible analogies between the vibrational spectrum of the filamentous phase which acquires the form $\omega^2 = a q_z^4 + c^2 q_\perp^2$ ($z$ is the direction along the chains axis) and the vortex crystal considered in [23, 25, 65].

Two Bose condensates in superfluid phase

Developing Ashcroft ideas [66], one of the authors of this Review (M. Yu. K.) [23, 64] advocated also an idea about the possibility of the unconventional superfluid (SF) state of metallic hydrogen at high pressures resembling superfluidity in neutron stars [67]. In this state formation and coexistence of two Bose condensates is possible, namely the Bose condensate of Cooper pairs in SC electron subsystem and Bose condensate of bi-proton pairs on one or neighboring chains or planes in SF ionic subsystem.

## 4. New results on metallic hydrogen and metallic hydrides

In this section we briefly announce our most recent results concerning the possibility to decrease pressure in complex metallic hydrides such as LaBH$_8$ [53]. In particular we will demonstrate the plot for the dependence of critical pressure from temperature with the negative derivative of $dT_C/dP < 0$ for the pressure range from 60 to 100 GPa in the triple hydride LaBH$_8$.

The results of calculation of $T_C$ for the triple hydride LaBH$_8$, performed for different pressures, are presented on Fig. 5, and for reduced density of states $N_0(E) = N(E)/N(0)$ – on Fig. 6. The calculations are performed in the pressure range from 60 to 100 GPa. Note that for smaller pressures this compound is probably unstable. We fulfilled the calculations utilizing two approaches. First approach is based on the solution of the "classical" system of Eliashberg equations without an account of the corrections to the chemical potential. In this method electron density of states does not enter in the equations (see e.g. [58]). Second one is based on the solution of the generalized system of Eliashberg equations which contains the corrections to the chemical potential [53].

Both methods showed decrease of $T_C$ with increase of pressure (starting from pressure values of 80 GPa). The results of [62] demonstrated decrease of $T_C$ for even in larger range of pressures. An account of the correction to the chemical potential of electrons yields increase of $T_C$ for all the pressures from 60 to 100 GPa. This increase is related to the presence of a smeared peak in electron density of states localized on the distance of the order of Debye energy from the Fermi level (see Fig.6). Note that behavior of $N_0(E)$ for energies larger than $\omega_D$ practically has no effect on the values of $T_C$. When we decrease pressure, we observe the growth of the peak in the density of states and correspondingly an increase of the critical temperature $T_C$.

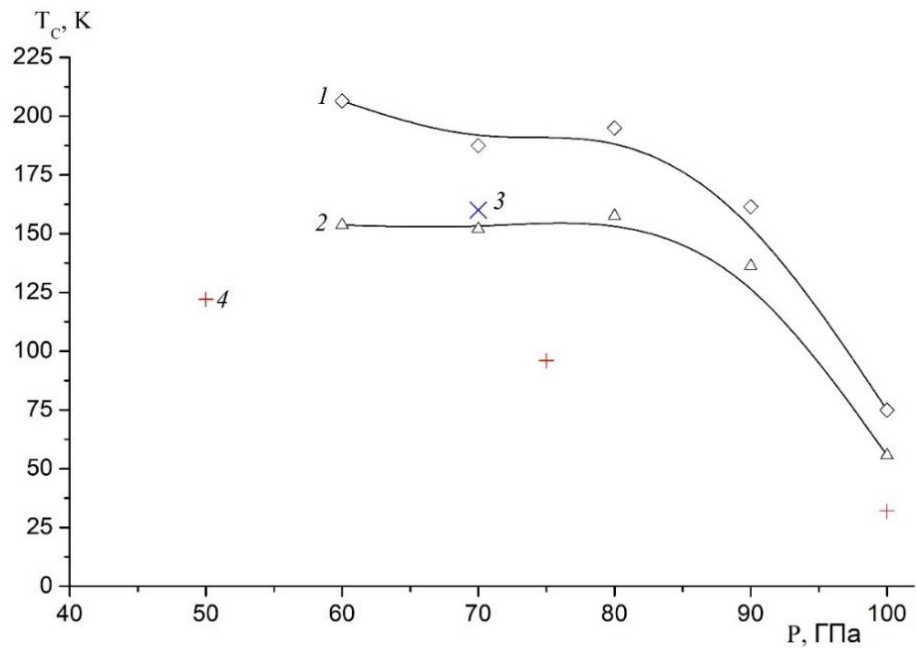

**Fig. 5.** Dependence of the critical temperature $T_C$ on pressure in LaBH$_8$, calculated utilizing different approaches: *1* – obtained from the solution of the generalized system of Eliashberg equations accounting for the corrections to the chemical potential [53]; *2* – from Eliashberg equations without corrections to the chemical potential; *3* – from the results of [68]; *4* – from the results of [58].

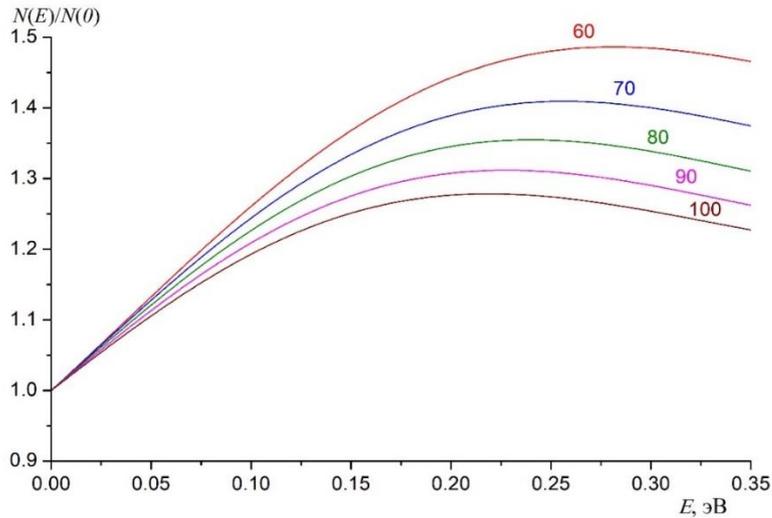

**Fig. 6.** Reduced density of states $N_0(E) = N(E)/N(0)$ for LaBH$_8$, calculated at different pressures. The pressure values (in GPa) are indicated close to the corresponding curves on the figure. Zero of energy corresponds to the Fermi level [53].

## 5. Conclusion

In the Review we discussed different mechanisms of superconductivity and inhomogeneous states in electron systems with attraction. In particular we considered SC mechanisms related to the BCS-BEC crossover and formation of local pairs for the bismuth oxides BaKBiO and two-dimensional low density electron systems with local (Hubbard) attraction in clean and dirty case. In these systems SC critical temperatures usually are not very high (in bismuth oxides they reach 30 K). At the same time, very interesting droplet structures (or clusters) of bosonic insulator are formed in these systems. When we increase the carrier density, the percolation insulator-superconductor phase transition is realized. As a result, we proceed from bosonic insulator to another interesting phase of bosonic metal.

Metallic hydrogen and metallic hydrides with the highest critical temperatures of the order of 200-250 K presumably are described by conventional Eliashberg mechanism of SC for systems with strong electron-phonon interaction. However, due to a number of reasons, metallic hydrogen and metallic hydrides are very interesting systems as well. Firstly, long-living metastable phases serving as local minima of the thermodynamic potential at low pressures can be formed in the normal state of these systems. Secondly two Bose condensates of Cooper pairs in electron subsystem and bi-proton pairs in ionic subsystem can describe the superfluid state.

Let us stress that normal state of low-dimensional metastable phases of metallic hydrogen and in particular of the filamentous and planar phase has important similarities with quantum (vortex) crystals.

Very important problem which should be solved on the way to the practical implementation of the room-temperature SC is the problem to increase the critical temperature when we decrease the pressure. We demonstrated the possibility of this effect for the triple hydride LaBH$_8$.

Finally, one more very interesting possibility for future investigations can be connected with the substitution of phonon modes by acoustic plasmon modes (predicted some time ago by P. Nozieres and D. Pines [69-70]). Acoustic plasmons could serve as intermediate bosons for different mechanisms of SC in the Random phase approximation (RPA) for electron plasma [1-3, 61, 71-72] including Frohlich mechanism [71].

**Acknowledgements**

M. Yu. K., R. Sh. I. and A.V. K. acknowledge the support of the HSE University (Program of Basic Research). The authors are grateful to K.I. Kugel and A. V. Kuznetsov for stimulating discussions.